\documentclass[12pt]{iopart}

\usepackage{graphicx}
\usepackage{dcolumn}
\usepackage{color}
\usepackage{bm}

\begin{document}

\title{Detection of charge motion in a non-metallic silicon isolated double quantum dot}

\author{T.~Ferrus, A.~Rossi, M.~Tanner, G.~Podd, P.~Chapman, and D.~A.~Williams}

\address{Hitachi Cambridge Laboratory, J. J. Thomson Avenue, CB3 0HE, Cambridge, United Kingdom}
\ead{taf25@cam.ac.uk}

\begin{abstract}

As semiconductor device dimensions are reduced to the nanometer scale, effects of high defect density surfaces on the transport properties become important to the extent that the metallic character that prevails in large and highly doped structures is lost and the use of quantum dots for charge sensing becomes complex. Here we have investigated the mechanism behind the detection of electron motion inside an electrically isolated double quantum dot that is capacitively coupled to a single electron transistor, both fabricated from highly phosphorous doped silicon wafers. Despite, the absence of a direct charge transfer between the detector and the double dot structure, an efficient detection is obtained. In particular, unusually large Coulomb peak shifts in gate voltage are observed. Results are explained in terms of charge rearrangement and the presence of inelastic cotunneling via states at the periphery of the single electron transistor dot.

\end{abstract}

\maketitle

\section{Introduction}

Early experiments on charge quantization in tunnel junctions \cite{tunnel junction} have initiated the development of single electron tunneling devices. However, this is not before the theoretical work of Averin on Coulomb oscillations (COs) \cite{Averin1} and the improvement of electron beam lithography that single electron transistors (SETs) were fabricated \cite{SET}. Their ability to detect the transfer of a single electron via Coulomb blockade (CB) with high efficiency made them usable in many architectures as charge pumps \cite{pump}, single electron memories \cite{SEMem}, quantum cellular automata \cite{automata} or in quantum computation \cite{quantum computing} as a charge detector. For the ease of operation, as well as for reliability, these nanometer-scale transistors are designed to have a metallic character.
In particular, irregularities in the confining potential are screened so that the internal electronic structure is well represented by energy levels whose separation is defined by the dot diameter. In such \textit{metallic} dots, localized states play a negligible role in the transport and sequential tunneling events are well predicted by the orthodox theory of CB \cite{Averin2} which relates the electron dynamics to bi-dimensional electron gas systems.

The \textit{metallic} property of the SET could be controlled by various methods, amongst which the most commonly implemented are the use of a center metal island and metal oxide tunnel barriers, as in Al/Al$_2$O$_3$/Al SETs \cite{AlSET}, the patterning of depletion gates in GaAs-based detectors \cite{split gate} or the realization of a metal-oxide-semiconductor (MOS) structure as in fin field effect transistor devices (FinFET) \cite{FinFETs} or in nanowire-based SETs \cite{Fujiwara}. Nevertheless, the sensitivity of \textit{metallic} SETs is limited by the 1/$f$ noise due to charge fluctuations at the dielectric-metal interfaces (Al/Al$_2$O$_3$) or trap charges (PB1 centers) at the Si/SiO$_2$ (100) interface in MOS structures. This explains that high frequency measurement or synchronous detection by two independent SETs may need to be performed to reduce the noise level \cite{RFSET}.

If a MOS structure is not used, silicon has to be doped to a high level and constrictions have to be patterned to help control the location of the formation of tunnel barriers \cite{SiSET}. In such a small system, randomness in dopant distribution and surface roughness could be responsible for device instabilities at low temperatures. Indeed, localized states that may be present, especially at the edge of the device, are sensitive to variation in the electrostatic potential. The problem of knowing the specific local potential within a non-bulk semiconductor nanostructure remains a challenge for many types of device, so the recognition of process-dependent characteristics is useful for obtaining indirect information about that potential.

When used as detectors, SETs are generally electrically coupled to a device made of a single or a double dot that is directly or indirectly connected to source and drain contacts  \cite{dots}. All these architectures are generally thought to be efficient in terms of charge detection owing to the strong coupling between the tunneling electron and the detector. Nonetheless the back-action from the detector to the device is substantial and electrical connections to the device and/or between the SET and the device are a non-negligible source of noise. On the contrary, the use of a geometrically isolated structure but capacitively coupled to the detector improves the electrical isolation but makes the detection more difficult. However, if these localized states at the detector edge could be controlled by geometric or electrostatic means, then they may be used to enhance the detection or to detect weaker effects.

In this article, we investigate a structure made of highly phosphorous doped silicon and comprising an isolated double quantum dot (IDQD) with a capacitively coupled single electron transistor. In such a device, the edge states occupancy could be modified by geometrical or electrostatic means, so that the coupling strength between the detector and the double dot structure could be engineered, leading to an efficient detection of electron motion in the IDQD. Observation of Coulomb peak shifts in gate voltage as well as cotunneling give an insight into the complex electron dynamics in this system and the important role played by edge localized states.

We first briefly introduce the device structure and the measurement setup. In section 3, we describe the observation of IDQD lines in a gate stability diagram and present a first attempt at simulating the experimental data and at describing the charge states dependence on gate voltages in the IDQD-SET system. In section 4, we discuss the detection mechanism in the SET by reviewing classical capacitance-based models and a trap-assisted tunneling model. This paper concludes with a summary in section 5.

\section{Devices and measurement setup}

The devices are fabricated from a silicon-on-insulator (SOI) wafer with a 45 nm-thick silicon layer, doped with phosphorous at a density of $\sim 2.9\,10^{19}$\,cm$^{-3}$. High resolution electron beam lithography and reactive ion etching were used to pattern a single dot of diameter $\sim$75\,nm with 30-nm width tunnel barriers for the detector as well as an isolated double dot of $\sim$75\,nm diameter. After oxidation the silicon dots were reduced to 60\,nm diameter with a lateral oxide thickness of 17\,nm. The devices are controlled by three in-plane gates that are formed from the same SOI layer (Fig. 1), one controlling the SET and the two others the double dot. The silicon substrate was connected to the ground. A custom low temperature complementary metal-oxide-semiconductor circuit (LTCMOS) is used to provide the various voltages to the device and to measure the SET current through a charge integrator. This arrangement enabled an efficient noise gain suppression by using shorter cabling between the measurement circuit and the device as well as sensitive and fast current detection compared to conventional measurements using room temperature source-measure units \cite{Hasko}. All lines were filtered by single stage low-pass resistance-inductance-capacitor filters with a cut-off of about 80\,kHz to suppress electrical noise and minimize the electron heating. The device and the filters are protected from radiated electrical noise by Faraday cages. Both the device and the LTCMOS were kept at 4.2\,K by immersing the probe into liquid helium. Several devices were processed identically with similar dimensions, some from different wafers. All showed similar characteristics and behavior at low temperature or during thermal cycling.

\begin{figure}
\begin{center}
\includegraphics[width=85mm, bb=0 0 300 200]{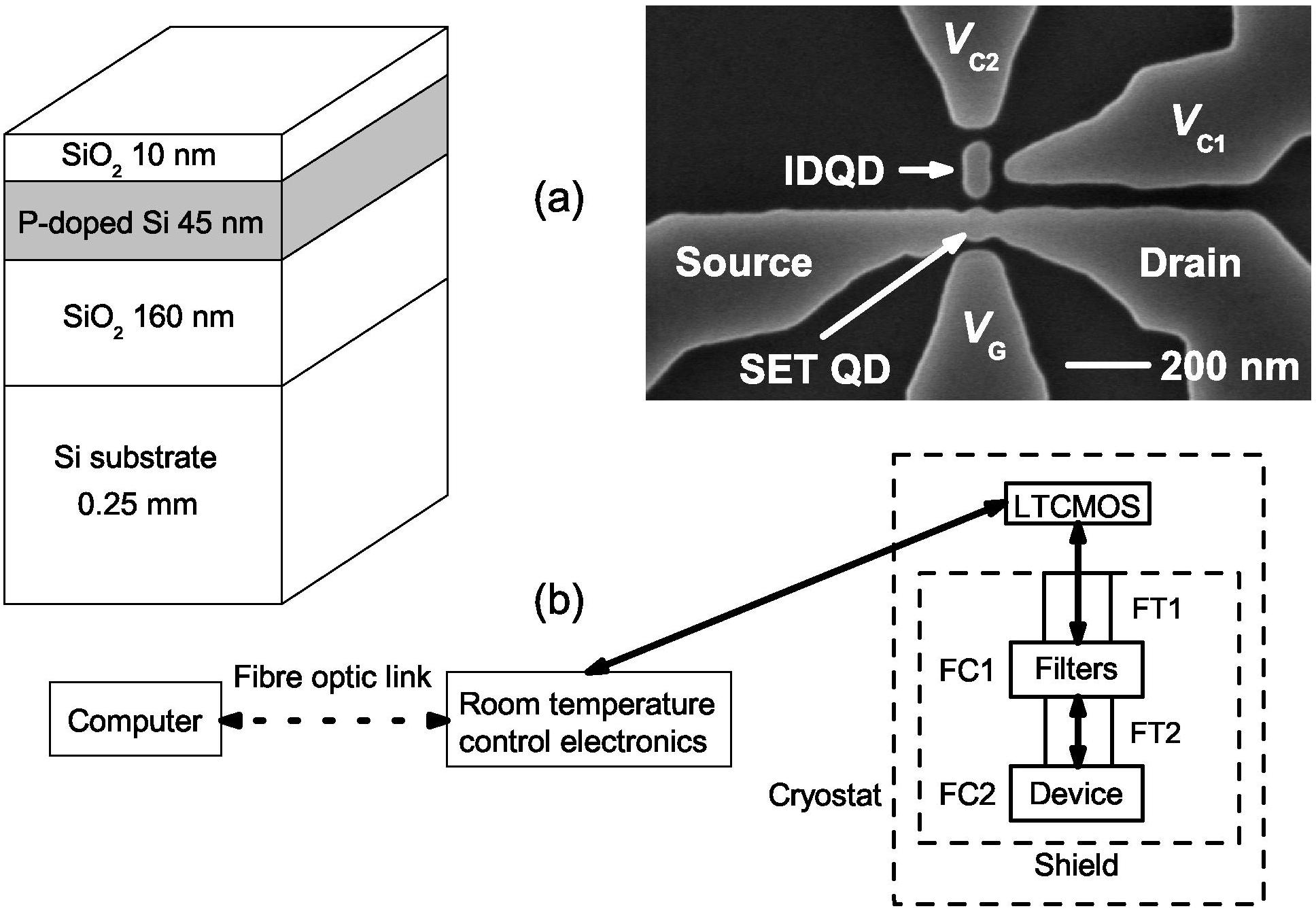}
\end{center}
\caption{\label{fig:figure1} a) Device structure and scanning electron microscope (SEM) image of the double dot with the nearby SET after oxidation. The upper IDQD gate was not fully functional and was grounded. b) Schematic representation of the measurement setup using the LTCMOS, Faraday cages (FC1 and FC2) and feed-through ferrite beads (FT1 and FT2).}
\end{figure}

\section{IDQD charge states detection}

In order to get an insight into the mechanism of charge motion detection in the IDQD, we have measured the dependence of the SET source-drain current $I_{\textup{\tiny{SD}}}$ on the SET gate ($V_{\textup{\tiny{G}}}$) and IDQD gate voltages ($V_{\textup{\tiny{C1}}}$ and $V_{\textup{\tiny{C2}}}$)(Fig 2). In the interest of simplifying the data analysis, $V_{\textup{\tiny{C2}}}$ (Fig. 1a) was grounded in most experiments. If not, its voltage was kept within the range $-0.9\,$V$\,< V_{\textup{\tiny{C2}}} < 0.5$\,V to avoid gate leakage. The obtained gate stability diagram $I_{\textup{\tiny{SD}}} \left(V_{\textup{\tiny{G}}}, V_{\textup{\tiny{C1}}} \right)$ clearly shows the existence of regions of gate voltages with anomalously low or high current, that are associated with the presence of localized states in the device, as well as the usual COs (SET lines). Both features are detailed in appendixes A and B, respectively. Additional features associated with the presence of the IDQD are also observed and discussed in the following sections.

\begin{figure}
\begin{center}
\includegraphics[width=85mm, bb=0 0 300 200]{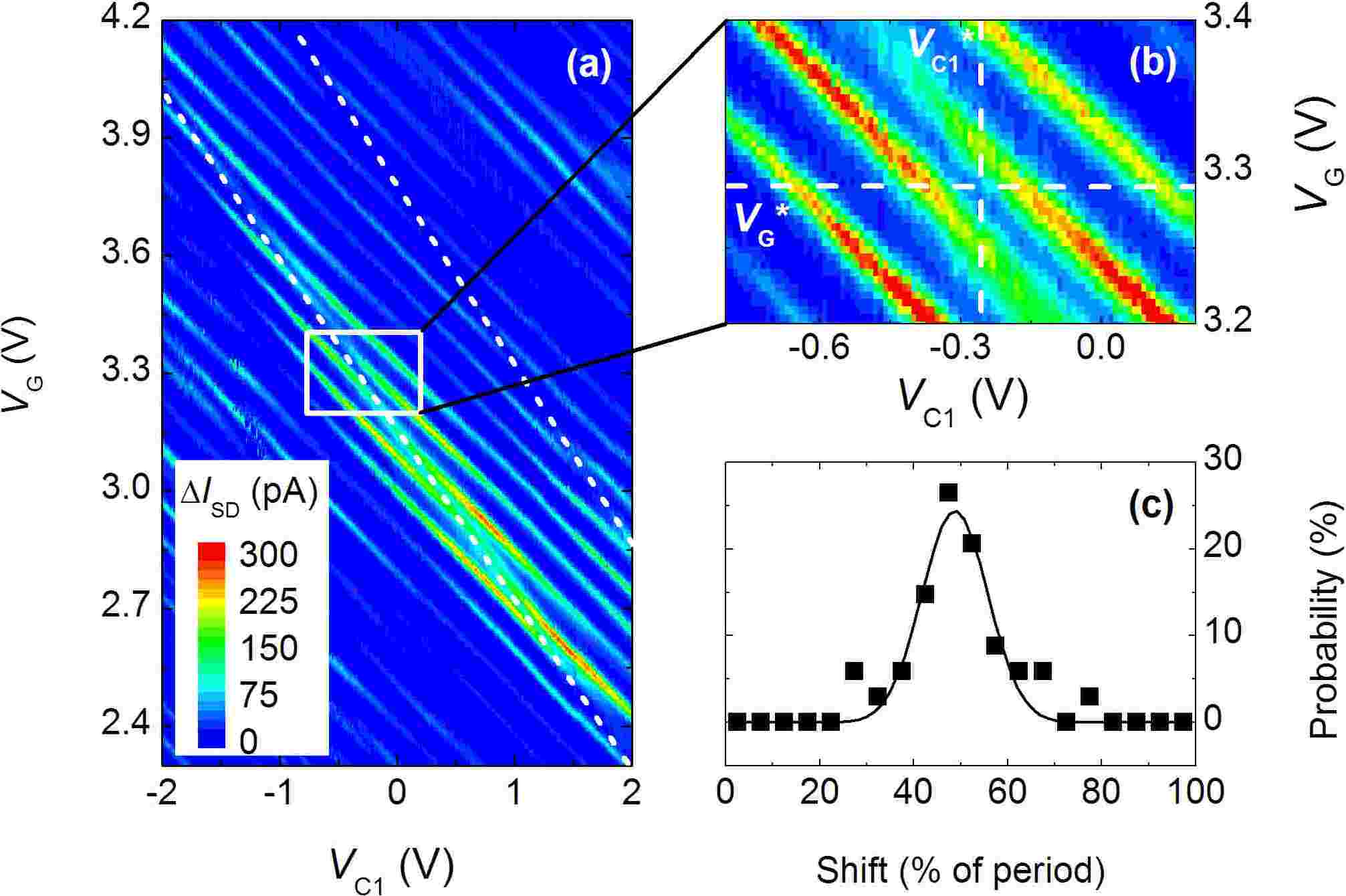}
\end{center}
\caption{\label{fig:figure2} a) Gate stability diagram with CO and IDQD lines (dotted lines). b) Shift as seen at the intersection between SET and IDQD lines. c) Distribution of the value of shift as a percentage of the CO period taken from different devices, cooldown and gate voltages. Most probable value is 49 $\%$.}
\end{figure}

\subsection{IDQD lines}

A common observation in these devices is the presence of additional lines in the gate stability diagram with a slope steeper than the ones associated with the SET (Fig. 2a). When intersecting the COs, these lines either enhance the Coulomb peak conductivity at specific gate voltages ($V_{\textup{\tiny{G}}}^{\star}$ and $V_{\textup{\tiny{C1}}}^{\star}$), or locally split them into two branches ($V_{\textup{\tiny{C1}}}<V_{\textup{\tiny{C1}}}^{\star}$ and $V_{\textup{\tiny{C1}}}>V_{\textup{\tiny{C1}}}^{\star}$) separated by a region of low but finite conductivity. On each branch, the Coulomb peak is shifted from its normal position by about $25\pm 7\%$ of the CO period (Fig. 2c).

Unlike COs, the additional lines are not periodic. Their position and visibility are both strongly affected by thermal cycles, but their slope is almost constant ($\textup{d}V_{\textup{\tiny{G}}}/\textup{d}V_{\textup{\tiny{C1}}} \sim$0.43$\pm$ 0.05). They are always present in devices containing an IDQD but are never observed otherwise. Still, associating these lines with the IDQD requires a deeper analysis because of the absence of a direct electron transfer from the IDQD to the detector, and so, of the impossibility of observing hexagonal shapes in the gate stability diagram, as usually obtained in connected double quantum dots.

The presence of additional conductivity lines in stability diagrams have been referenced by a few authors. In GaAs/GaAlAs quantum dots \cite{Zhitenev}, they have been related to bound electrons at the periphery of the quantum dot when the device is tuned to be close to the delocalization-localization transition, e. g. a situation where both localized electrons at the dot boundary and delocalized electrons at the center of the island coexist. Further observations have been made by Gunther \textit{et al.} \cite{Gunther}, this time in a silicon metal-oxide-semiconductor field effect transistor (MOSFET)-like structure. In his paper, an alternative explanation is given and additional lines are expected to result from a modification of the quantum dot confinement that lifts the degeneracy of the quantized energy levels.

In both cases, these features originate from the specific internal electronic structure of the quantum dot. Although, regions of different localization strength do exist in our devices, there are a number of differences. Firstly, in Zhitenev's device, anticrossings are visible at the intersection between additional and SET lines. This means there is no discontinuity in the Coulomb peak positions in gate voltage, and so, no noticeable variation of the peak capacitance over the transition region. The peak position is not altered or only changed by a few percent, at either side of the crossing point. On the contrary, we do observe a clear discontinuity (or shift) in the Coulomb peak position in gate voltage as well as a strong variation in the conductivity at the crossing point. Secondly, the slope of the additional lines is, in our case, almost constant over a wide range of gate voltages, unlike Zhitenev or Gunther's observations. 

Sharps and clear shifts have been observed in doubly gated planar silicon MOS structure in the accumulation mode
 by Morello \textit{et al.} \cite{Morello}. In their experiments, the line slope $\textup{d}V_{\textup{\tiny{G}}}/\textup{d}V_{\textup{\tiny{C1}}}$ is very large, a distinctive feature of random telegraph signal or electron tunneling from an impurity outside but close to the detector edge into the detector itself. However, the edge of our SET has been electrically isolated by etching the surrounding silicon, so such a tunneling is unlikely.

\begin{figure}
\begin{center}
\includegraphics[width=85mm, bb= 0 0 300 200]{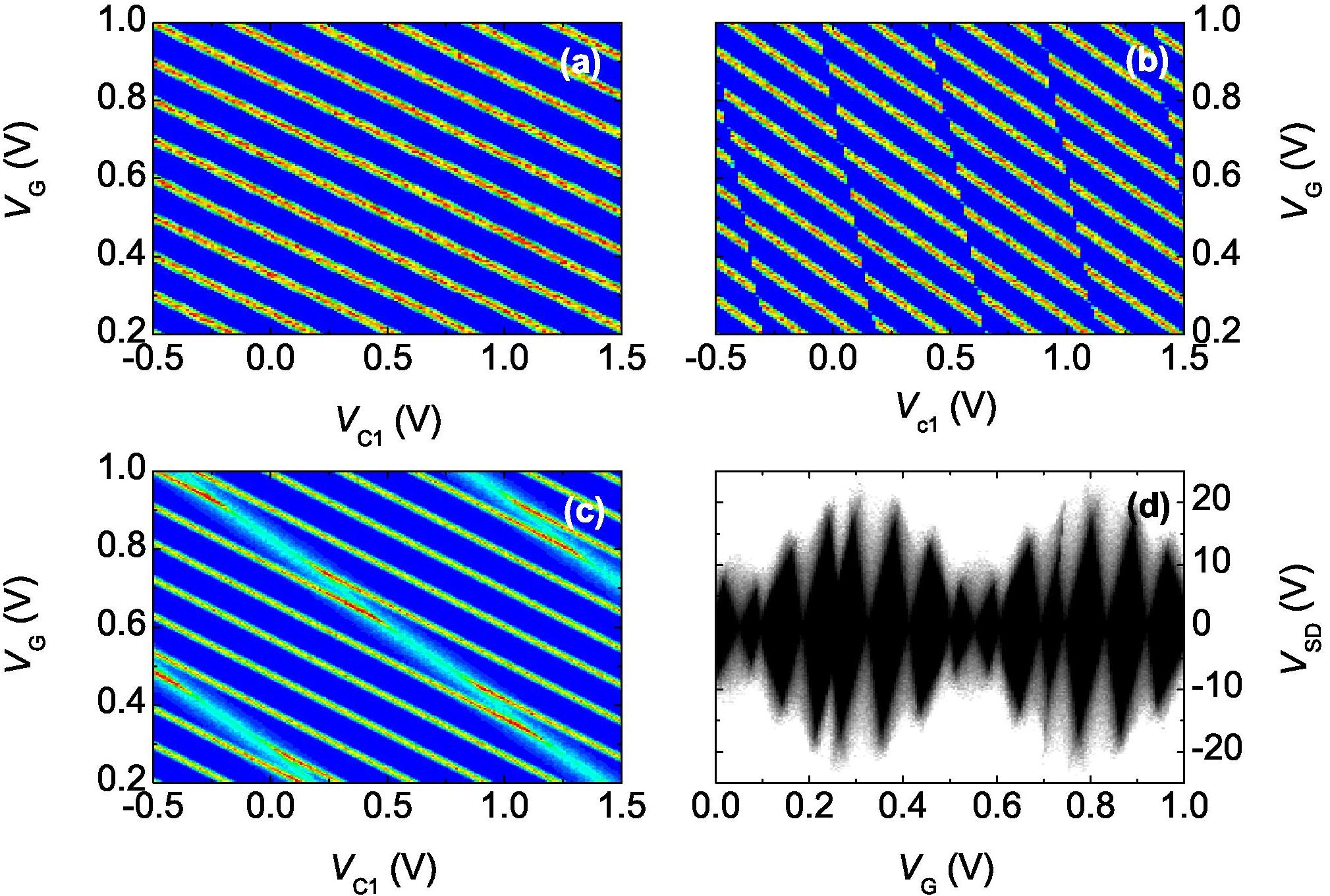}
\end{center}
\caption{\label{fig:figure3} Gate stability diagrams for the case of a SET alone (a), a SET connected to a single dot via a tunnel barrier (b), a SET with a capacitively coupled IDQD (c). (d) Coulomb diamonds obtained when adding a parallel conduction mechanism in the case of an SET-IDQD structure.}
\end{figure}

\subsection{IDQD coupling and trap assisted tunneling : simulations}

In order to establish the origin of the observed additional lines, we performed first principle calculations and simulations using SIMON 2.0, a single-electron circuit simulator based on Monte Carlo simulation \cite{SIMON}. Although neglecting the shape of the dots, the dopant distribution and many-body interaction, these simulations  give substantial indications on the structure responsible for the additional lines. 

Simulations that include a SET without an IDQD well reproduces the characteristic slope of the SET lines in the gate stability diagram (Fig. 3a). In simulations where a SET is connected via a tunnel barrier to a nearby trap charge, additional lines are obtained but with a slope d$V_{\textup{\tiny{G}}}/\textup{d}V_{\textup{\tiny{C1}}} \gg$1, similarly to the work of Morello \textit{et al.} (Fig. 3b) but unlike our experimental results.

However, the values for the SET and additional line slopes as well as for the shift were well reproduced when considering a SET that is capacitively coupled to a double dot system. Simulations reveal also that the IDQD is better modeled by a single but large dot where electrons are allowed to move between the lower and the upper IDQD dots via a tunnel barrier. This behavior is expected from the width of the tunnel barriers, the orientation of the IDQD with respect to the SET and gates, as well as from the size of the dots measured by SEM. The tunneling resistance between the two IDQD dots governs the visibility of the region around the shift whereas the intra-IDQD capacitance and the SET-IDQD capacitance influence the value of the shift itself.

Indeed, in such a structure, the electrostatic environment is modified by the different voltages applied to it with a strong interplay by impurities and charge reorganization (Appendix A). As a result, we expect capacitances to be gate voltage dependent. Such a variation in the inter-dot coupling strength is clearly revealed in figure 4.

\begin{figure}
\begin{center}
\includegraphics[width=85mm, bb=0 0 300 200]{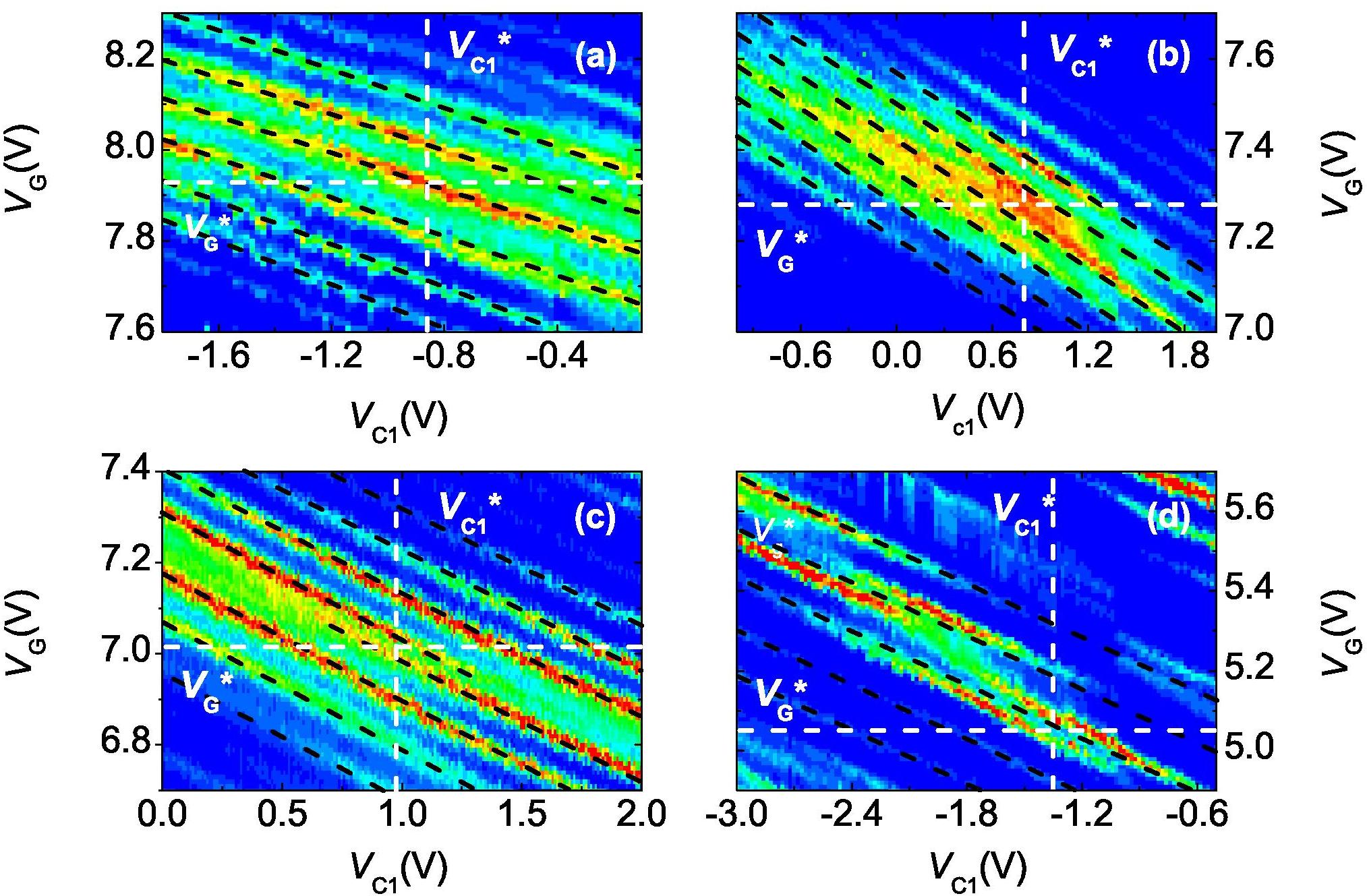}
\end{center}
\caption{\label{fig:figure4} Effects of the coupling strength between the IDQD and the SET on the SET current, from weak coupling (a) to strong coupling (d).}
\end{figure}

\begin{figure}
\begin{center}
\includegraphics[width=85mm,bb=0 0 300 200]{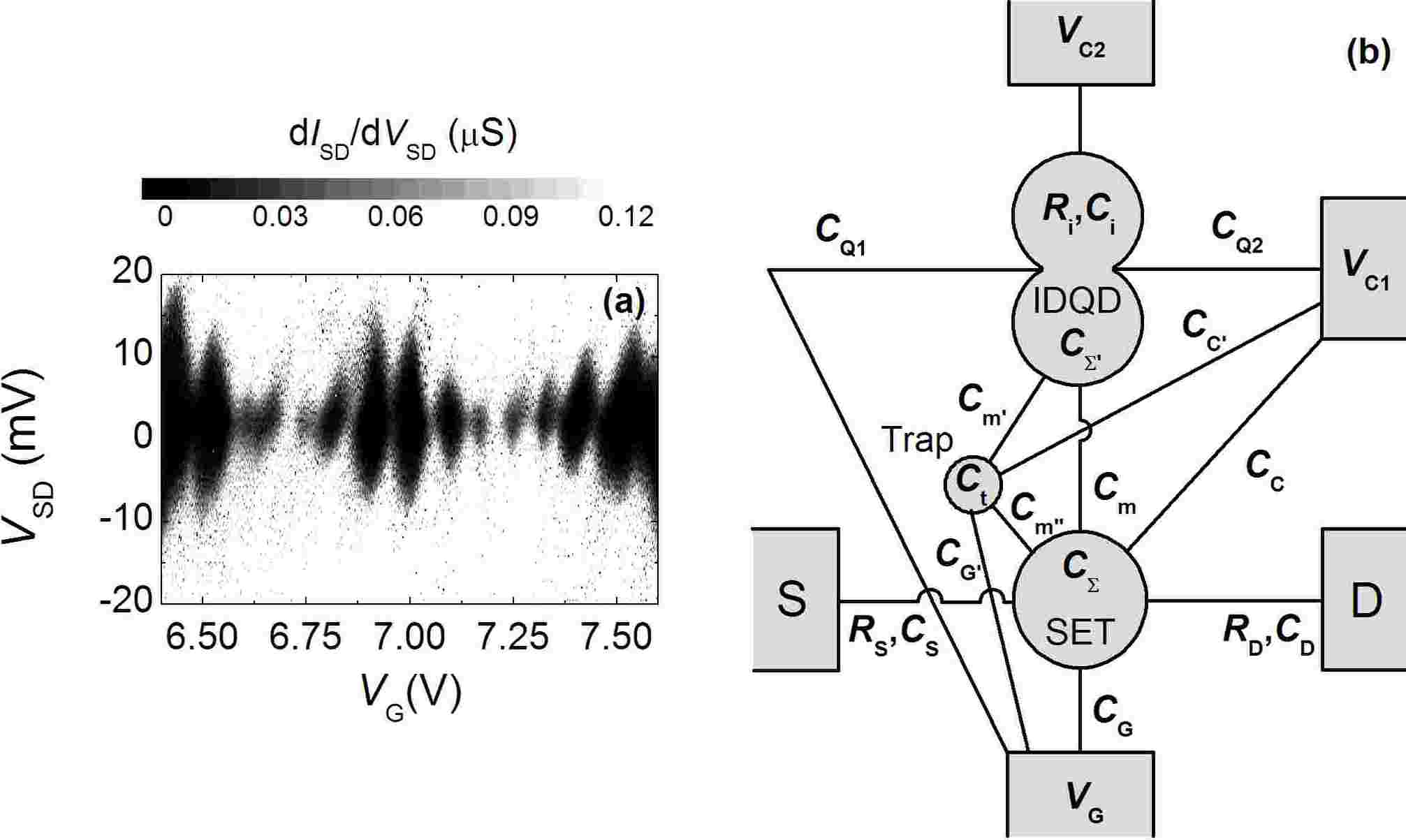}
\end{center}
\caption{\label{fig:figure5} a) Coulomb diamonds showing parallel dot like behavior. b) Schematic diagram for the capacitance and resistance circuit used for the simulations with a SET, an IDQD and a trap charge as well as the different gates and the source (S) and drain (D) contacts. For clarity not all capacitances are shown.}
\end{figure}

However, simulations failed to reproduce correctly a number of features in the current-voltage dependencies. The most striking is the shape of Coulomb diamonds. Experimentally charging energies are varying strongly with gate voltages and the extend of the Coulomb diamond along $V_{\textup{\tiny{SD}}}$ may be noticeably reduced in some regions (Fig. 5a). This effect can qualitatively be taken into account by adding a parallel tunneling process via a trap charge in the SET. The trap is simulated as a quantum dot connected to the SET and contacts via a tunnel barrier and capacitively coupled to the IDQD (Figs. 3c and d). Such a trapping mechanism also well explains the local increase in conductivity, controlled by the tunneling resistance between the contacts and the trap, and the broadening of the Coulomb peak at the shift position, controlled by the coupling capacitance between the trap and the IDQD (Fig. 3c). 

Figures 2a and 5a (Appendix B) were used for simulations and gate dependencies of the current were well fitted using the capacitances values listed in Table 1.

\begin{table}[ht]
\centering 
\begin{tabular}{c c c c} 
\hline\hline 
 $C_\textup{\tiny{G}}$   &   $C_\textup{\tiny{C}}$   &   $C_\textup{\tiny{Q1}}$   &   $C_\textup{\tiny{Q2}}$ \\ [0.5ex] 
\hline 
1.61  & 0.471  & 0.77  & 0.34 \\ [1ex] 
\hline\hline 
 $R_\textup{\tiny{S(D)}}$   &   $C_\textup{\tiny{S(D)}}$   &   $R_\textup{\tiny{i}}$   &   $C_\textup{\tiny{i}}$  \\ [0.5ex] 
\hline 
 133  &   8    & 200   &   0.4  \\ [1ex] 
\hline\hline 
 $C_\textup{\tiny{m}}$   &   $C_\textup{\tiny{m'}}$   &   $C_\textup{\tiny{m''}}$   &     \\ [0.5ex] 
\hline 
 0.8   &   0.2   &   0.2   &      \\ [1ex]
\hline \hline 
\end{tabular}
\caption{Values for capacitances (aF) and resistances (k$\Omega$) as defined in figure B2d.}
\end{table}

The capacitance values well agree with the ones obtained from FastCap 2.0 \cite{FastCap}, a software used for computing the self and mutual capacitances of a conductive tridimensional structure, whose dimensions were extracted from SEM imaging.

\subsection{Charge states in the IDQD-SET-trap structure}

Although simulations satisfactory explain experimental data and link the observation of additional lines to the presence of the IDQD, a few inconsistencies suggest a more sophisticated model may be needed. For example, in Fig. 4d, the SET lines have their conductivity strongly suppressed away from the shift position. Also, these lines are not deviating from their original position except at the intersection with the IDQD lines (local shift) contrary to the simulations where all SET lines are shifted together after the crossing (global shift). This might suggest a stronger influence of traps on the SET dynamics and the possibility of electron correlation.

Indeed, SIMON 2.0 allows electron transfer to the trap and to the SET at the same time without taking into account that a filled trap at the edge of the SET or close to the barrier may block further tunneling into the SET island due to Coulomb repulsion. This is because both the SET and the trap are considered as two separate structures. It is possible to improve the modeling by considering that the trap charges are located within the SET, so that the total charge of the SET and the trap has to be maintained constant in a blockade region. 

In a system made of a SET with a capacitively coupled IDQD, the dependence of the Coulomb peak position in gate voltage can approximately be determined without involving lengthly capacitance matrix calculations by noting that a change in the bias condition d$V$ is equivalent to a charge addition d$Q =C \textup{d}V$ in the structure and that a variation of charge d$Q'$ in the IDQD modifies the potential in the SET, leading to an effective charge addition in the detector of d$Q''=C_{\textup{\tiny{m}}}/C_{\Sigma'} \textup{d}Q'$, where $C_{\textup{\tiny{m}}}$ is the capacitance between the SET and the IDQD and $C_{\Sigma'}$ the total capacitance of the IDQD. Similarly to the SET lines, IDQD lines correspond to a tunneling of an electron between the two IDQD dots through the tunnel barrier separating them. By neglecting back-actions between the SET and the IDQD and following the previous comments on charge tunneling, the SET lines are given by

\begin{eqnarray}\label{eqn:equation2}
V_{\textup{\tiny{G}}} \approx -\frac{C_{\textup{\tiny{C}}}}{C_{\textup{\tiny{G}}}} V_{\textup{\tiny{C1}}} + \frac{e}{C_{\textup{\tiny{G}}}} N_{\textup{\tiny{S}}} + \frac{e C_{\textup{\tiny{m}}}}{C_{\Sigma'}C_{\textup{\tiny{G}}}} N_{\textup{\tiny{i}}} 
\end{eqnarray}

and the IDQD lines by

\begin{eqnarray}\label{eqn:equation3}
V_{\textup{\tiny{G}}} \approx -\frac{C_{\textup{\tiny{Q2}}}}{C_{\textup{\tiny{Q1}}}} V_{\textup{\tiny{C1}}} + \frac{e}{C_{\textup{\tiny{Q1}}}} N_{\textup{\tiny{i}}} + \frac{e C_{\textup{\tiny{m}}}}{C_{\Sigma}C_{\textup{\tiny{Q1}}}} N_{\textup{\tiny{S}}} 
\end{eqnarray}

where $e$ is the elementary charge, $N_{\textup{\tiny{S}}}$ and  $N_{\textup{\tiny{i}}}$ are the total number of electrons respectively in the SET and in the IDQD. The other notations are defined in Fig. 4d.

The first terms in Eqs. 1 and 2 give the corresponding slopes in the gate stability diagram, whereas the second give the period on the $V_{\textup{\tiny{G}}}$ axis and the third the shift in gate voltage relatively to \textit{ideal} position (without the capacitively coupled structure).

Because of the absence of electrical connections between the IDQD and the source or drain leads, $C_{\Sigma} \gg C_{\Sigma'}$, so that, experimentally, the IDQD lines are not significantly shifted when intersecting the SET lines. Unlike SET lines, we did not observe any periodicity in the IDQD lines experimentally. This may indicate that more complex mechanisms involving traps at the SET periphery and Coulomb interaction may be involved or, more directly, that capacitances, in particular $C_{\textup{\tiny{m}}}$, may depend on gate voltages.

By taking into account a trap charge at the edge of the SET, the shift of the SET peak position (in percent of the CO period) is

\begin{eqnarray}\label{eqn:equation4}
\gamma= \Delta N_{\textup{\tiny{t}}} \frac{C_{\textup{\tiny{m''}}}}{C_{\textup{\tiny{t}}}} + \Delta N_{\textup{\tiny{i}}} \left( \frac{C_{\textup{\tiny{m}}}}{C_{\Sigma'}} + \frac{C_{\textup{\tiny{m'}}} C_{\textup{\tiny{m''}}}}{C_{\Sigma'}C_{\textup{\tiny{t}}}} \right)
\end{eqnarray}

where $\Delta N_{\textup{\tiny{t}}}$ and $\Delta N_{\textup{\tiny{i}}}$ are respectively the change in the number of electrons in the trap and in the IDQD. 

Because the IDQD is electrically isolated from the rest of the device, its total charge, including localized and extended states, is conserved at all times. The SET conductivity can then only be affected if the coupling between the SET island and the IDQD quantum dots differs for the upper and lower dot, so that a charge displacement, or a significant change in the electron distribution in the IDQD is seen, by the SET, as an effective charge offset $\Delta N_{\textup{\tiny{i}}}$ in the IDQD structure. It should be noted that the perpendicular position of the IDQD relatively to the SET improves the difference in sensitivity of the two IDQD dots.

When following a Coulomb peak in the gate stability diagram but away from the crossing between the SET and IDQD lines, the charge is constant in the IDQD and there is a single tunnel event in the SET-trap system. Thus $\Delta N_{\textup{\tiny{i}}} = 0$ and $\Delta N_{\textup{\tiny{t}}}+ \Delta N_{\textup{\tiny{S}}}=1$, with $\Delta N_{\textup{\tiny{S}}}$ the variation in the electron number in the SET. In this case, either the electron tunnels to the SET island and $\gamma=0$ (usual tunneling) or it tunnels to the trap and $\gamma=C_{\textup{\tiny{m''}}} / C_{\textup{\tiny{t}}}$ (trap assisted tunneling). Because $\gamma$ is independent of $N_{\textup{\tiny{i}}}$, there is no global shift in the position of the SET line. However, at the shift position $\Delta N_{\textup{\tiny{i}}} = 1$ so that the SET lines are shifted by $\sim \pm C_{\textup{\tiny{m}}}/ 2 C_{\Sigma'}$ depending on the position of $V_{\textup{\tiny{G}}}$ and $V_{\textup{\tiny{C1}}}$ relatively to $V_{\textup{\tiny{G}}}^{\star}$ and $V_{\textup{\tiny{C1}}}^{\star}$. Finally, following the IDQD line ($\Delta N_{\textup{\tiny{i}}}=1$ ) but in the blockade region ($\Delta N_{\textup{\tiny{t}}}+ \Delta N_{\textup{\tiny{S}}}=0$), the increase of conductivity may result from cotunneling via the trap states (one electron tunneling to the trap and one electron leaving the SET island for charge conservation). In this case, cotunneling is expected to be inelastic because of the activation energy of the trap and an estimate value for the energy difference has to be investigated (Appendix C).

\section{Detection mechanism}

\subsection{Classical models}

Theoretical predictions that are based on purely static modeling, such as image charges, all predict a value for the shift close to 4$\,\%$ of the CO period \cite{Lee,Image}. The model is based on the existence of interfaces and materials of different permittivity and neglects electron dynamics. It explains the experimentally measured value for the shift in the case of a mobile charge in nanocrystalline silicon quantum dots where detection is made by a multiple-gate single-electron transistor \cite{shift} as well as in other \textit{metallic}-like systems. This elementary model is also well suitable for capacitively coupled but weakly interacting systems \cite{Fujisawa} or for a similar system where the double quantum dot is parallel to the detector and connected to an electron reservoir \cite{Rossi}.

This model is clearly unsuitable in our case for reasons discussed in the previous sections, in particular the presence of a certain level of localization, the importance of tunneling via traps, as well as and the isolated character of the IDQD. Indeed if the IDQD was connected to a lead, then the effects of trapping and charge reorganization would have been significantly screened due to the continuous interaction between electrons in the dots and those in the reservoir.

More sophisticated capacitance modeling, like SIMON's provided significant information on the dynamics of the system but may not be entirely appropriate for a doped isolated structure. Indeed traps are still simulated as \textit{metallic} dots and electron interaction is neglected. This explains the inconsistency in the value of $C_{\textup{\tiny{m''}}}$ which is expected to be large due to the strong coupling between trap and the SET island, although providing a reasonable agreement with experimental data. Indeed, capacitance is irrelevant for such a localized state. Nevertheless, it is important to assess the extent of its validity. In such a model, quantum dot structures do not need to be explicitly electrically connected except when dealing with charge conservation. In particular, the voltage shifts between the SET lines can always be associated with a measure of the electrostatic coupling between the dots, including the case of an isolated structure like the IDQD \cite{Hofmann}. The equivalent coupling capacitance between the IDQD and the SET is then given by

\begin{eqnarray}\label{eqn:equation5}
C_{\textup{\tiny{m}}} = \gamma C_{\Sigma'}
\end{eqnarray}

where $\gamma$ is the Coulomb peak shift in percent of the CO period.

From Sec. 3.3, we also have

\begin{eqnarray}\label{eqn:equation6}
{\it{\Delta}}=\frac{e}{C_{\textup{\tiny{Q1}}}}
\end{eqnarray}

\begin{eqnarray}\label{eqn:equation7}
S=-\frac{C_{\textup{\tiny{Q2}}}}{C_{\textup{\tiny{Q1}}}}
\end{eqnarray}

where ${\it{\Delta}}$ is the separation between successive IDQD lines in gate voltage and $S$ their slope in the ($V_g, V_c$) gate dependency diagram.

From these relations, we obtain

\begin{eqnarray}\label{eqn:equation8}
C_{\textup{\tiny{m}}} = \frac{\gamma}{1-\gamma} \left[\frac{ 2 e}{\it{\Delta}} \left( 1-S \right) + C_{\textup{\tiny{i}}}\right]
\end{eqnarray}

$\gamma$ and $S$ being almost constant, the non-periodicity of the IDQD lines in gate voltage clearly shows that one has to consider a dependence of $C_{\textup{\tiny{m}}}$ on gate voltage.

By taking ${\it{\Delta}}=0.207$\,V as the minimum observable value and $C_{\textup{\tiny{i}}}=0.4$\,aF, we obtain $C_{\textup{\tiny{Q2}}} \sim 0.33$\,aF, $C_{\textup{\tiny{Q1}}} \sim 0.77$ aF, $C_{\Sigma'} \sim 2.34$\,aF and $C_{\textup{\tiny{m}}} \sim 0.84$\,aF. However, for the maximum experimentally measured value, ${\it{\Delta}}=0.453$\,V, we have $C_{\textup{\tiny{Q2}}} \sim 0.15$\,aF, $C_{\textup{\tiny{Q1}}} \sim 0.35$ aF, $C_{\Sigma'} \sim 1.91$\,aF and $C_{\textup{\tiny{m}}} \sim 0.61$\,aF. Good agreement is obtained between these experimental values and theoretical calculations if considering an effective relative permittivity of 2.1. This low value reflects that the SET and the IDQD are surrounded by trenches filled, at low temperature, with liquid helium with permittivity close to 1. Although giving satisfactory results, this approach is an equivalent model for which the capacitance calculation is adjusted to an experimental observed value for the shift. Nevertheless, this model has the advantage to show that the present experimental results cannot be understood without considering a dependence of the capacitance in gate voltage and possibly electron dynamics. This, in turns, confirms again the presence of a variable electrostatic environment, which is easily modeled by localized states at the edge of the SET, as previously discussed. 
It shall be noted that such a variable coupling is expected to strongly influence the SET gate capacitance $C_{\textup{\tiny{G}}}$ as well. On the contrary, most of the variation in the gate periodicity, with the exception of Coulomb peak shift region, can be attributed to a variation in the value of the excited states energy so that $C_{\textup{\tiny{G}}}$ is not varying more than $10 \%$ of its mean value. Indeed, the presence of a source and drain contact, provide an equilibrium mechanism to the displacement of electrons inside the SET island, so that the variation in the electron distribution is compensated by electron entering or leaving the SET island. This situation is very different in the double dot because of its insulation from the electrical environment, and such a compensation cannot take place. This coupling is generally weak, except when the electrostatic arrangement is favourable to a charge motion in the IDQD and its detection by cotunneling effect in the SET island. This also suggests the possibility of charge rearrangement in both the IDQD and the SET \cite{arrangement}. However, this model does not provide any information on the energies involved.

\subsection{Charge ring model}

Previous observations suggest that a significant proportion of traps and localized states are found at the periphery of the SET island (Appendix A). Their distribution and the traps occupation number can eventually be controlled by fabricating a backgate and adjusting its voltage during the device cooldown. Such a method has already been applied successfully on metal-oxide-semiconductor (MOS) based devices \cite{Pepper}. The presence of such a distribution of charge in a ring-like shape around a quantum dot was suggested by Zhitenev \textit{et} al \cite{Zhitenev} and studied by Rudin \textit{et} al \cite{Rudin} in floating gate transistors. Following the same method but adapting it to the present geometry, an electron getting trapped at the periphery of the SET island leads to a shift in gate voltage, from its normal position, given by :

\begin{eqnarray}\label{eqn:equation9}
\delta V_{\textup{\tiny{G}}} = \frac{\textup{e}}{{\pi}^2 {\epsilon}_0 {\epsilon}_r \left( r-R \right)} \textup{K} \left[ \frac{-4R r}{\left( R-r \right) ^2} \right]
\end{eqnarray}

where $R$ is the SET radius including the silicon and oxide region, $r$ is the distance from the center of the SET island to the edge of the SET gate, ${\epsilon}_r \sim 2.1$ the effective permittivity of the trench and $\textup{K}\left(x\right)$ the complete elliptic integral of first kind.

Supposing the trap is at the vicinity of the Si-SiO$_2$ interface, we take $R \sim 30$\,nm and $r \sim 60$\,nm (Appendix B) and estimate $\delta V_{\textup{\tiny{G}}} \sim$ 25\,mV. This leads to a shift $2 \delta V_{\textup{\tiny{G}}} / \Delta V_{\textup{\tiny{G}}}$ of about 51 $\%$ in excellent agreement with the experimental value. The large observed value is mostly due to the fact the side gate and the SET island are electrically isolated by a trench. This may explain why such a large shift has never observed in conventional gated devices where the permittivity is a factor 10 higher.

Within this model, the energy necessary to trap one electron at the periphery is given by the mean single particle level spacing ${\it{\Delta}}_1$. This is confirmed by the presence of a the central Coulomb diamond (I) in figure C2a which charging energy is about 1.8\,meV and corresponds to the energy difference between the trapping of an extra electron and the removal of a trapped electron. It is interesting to notice that, when entering the dot the electron can tunnel into two possible states at $E_{\textup{\tiny{C}}}$ and $E_{\textup{\tiny{C}}} + {\it{\Delta}}_1$. However, the $E_{\textup{\tiny{C}}} + {\it{\Delta}}_1$ state has a stronger coupling to the edge states because the potential energy at the edge of the dot ($m^{\star} {\omega_0}^2 R^2/2$) corresponds to ${\it{\Delta}}_1$, by definition. It is also the point where the kinetic energy is zero, so where the localized states are most likely to be. Such a coupling favors charge reorganization at a minimum cost within ${\textup{e}}^2 / \left(4 \pi \epsilon r \right)\textup{e}^{-\lambda r}$, where $\lambda$ is the Thomas-Fermi screening length, if taking into account electron screening. Although the energy cost is higher via this process, the overall cost is lowered due to electron rearrangement so that the real cost in energy between the two processes, via normal tunneling and via cotunneling, is $\Delta E = {\it{\Delta}}_1-{\textup{e}}^2 / \left(4 \pi \epsilon r \right)\textup{e}^{-\lambda r}$. The maximum value for $r$ is the dot diameter corresponding to one electron being trapped at the source and one released at the drain so that $\Delta E = -0.2$\,meV but values up to -2.3\,meV could be obtained depending on the configuration.

\subsection{Transport mechanism}

It should be noted that the electron trapping mechanism at the periphery of the SET dot as described previously, as well as the value of the gate voltage shift given by Eq. 8 do not explicitly reference the presence of an IDQD, nor an electron displacement in the IDQD. However, in the absence of an isolated structure, the effect is expected to be random, rare and generally hidden by direct electron tunneling since it is a second-order tunneling process. 

The presence of the IDQD modifies the transport in different ways. When no electrons are transferred in the IDQD, but the IDQD is polarized, the effective charge in the lower dot either suppresses the SET tunneling current by populating the SET periphery states, e.g reducing the effective dot size or enhances the current in the blockade regime by displacing the electrons towards the inner region of the SET and minimizing scattering from localized electrons (Appendix B.2). In addition, since the IDQD and the SET are made from the same material with similar dimensions, the displacement of an electron in the IDQD induces both a charge reorganization in the IDQD and a modification of the electric field at the upper of the lower IDQD dot. Due to capacitance coupling and under the appropriate gate voltages, the electrostatic potential is modified in the SET island with the strongest effect expected to happen at the periphery of the SET. This allows the trap occupation number to be modified at the edges of the SET island at no energy cost, making cotunneling an efficient and dominant process (Sec. 4. 2).

Thus, the presence of a charge movement in an isolated structure provides a mechanism for the suppression of direct tunneling and the enhancement of trap assisted cotunneling in the blockade regime, at specific and reproducible combinations of gate voltages, making the effect more visible and controlled. As a consequence, the direct observation of Coulomb peak shifts gives indication on the effective polarization of the IDQD, $V_{\textup{\tiny{G}}}^{\star}$ and $V_{\textup{\tiny{C1}}}^{\star}$ indicating the gate positions of the IDQD state degeneracy.

\section{Conclusions}

We have shown that a highly phosphorous doped silicon single electron transistor can efficiently detect charge movement in a nearby but electrically isolated double dot despite the absence of a direct electron transfer between the two structures. The presence of localized states at the periphery of the SET dot and the ability for the system to proceed to charge rearrangement allow inelastic cotunneling to be an efficient conduction mechanism. In particular, the most noticeable effect is the presence of significant Coulomb peak shifts in gate voltage. In such devices, electron dynamics are complex. Nevertheless, the glass-like behavior of the systems allows charge reorganization to take place and detection to remain efficient. These results thus extend the possibility of realizing and detecting charge qubits in non-\textit{metallic} devices.

This work was partly supported by Special Coordination Funds for Promoting Science and Technology in Japan.

\appendix

\section{Impurities and localization}

\subsection{Conductivity background}

The most noticeable feature in figure A1 is the existence of a large conductivity background on top of which lie the usual COs. It is aperiodic and only reproducible within a single thermal cycle. This discards a purely electrostatic influence from the gate voltages or the creation of built-in potentials inside the structure. On the contrary, this suggests the presence of charging effects and the probable influence of localized states. This effect was confirmed on all devices with or without the double dot structure.

The conductivity background is gate voltage dependent, with regions where the SET current is anomalously suppressed and others where it is significantly enhanced. These features are found along lines with a slope $\textup{d}V_{\textup{\tiny{G}}}/\textup{d}V_{\textup{\tiny{C1}}} \sim$0.23$\pm 0.06$ in the gate stability diagram, a value close to the one observed for the COs. Therefore, it is likely that they originate from the main SET island. 

Indeed, for a device containing traps, a thermal cycle allows electrons to be redistributed amongst localizing centers. So the electrostatic potential due to these charges is likely to be modified each time the device is thermally cycled (Fig. A1a as compared with Fig. A1b). 

Because the IDQD is made of the same material than the SET, such a charge reorganization is also expected in the double dot structure. However, owing to its isolation from the rest of the device, it is difficult to probe separately the effect in the IDQD.

\begin{figure}
\begin{center}
\includegraphics[width=85mm,bb=0 0 300 200]{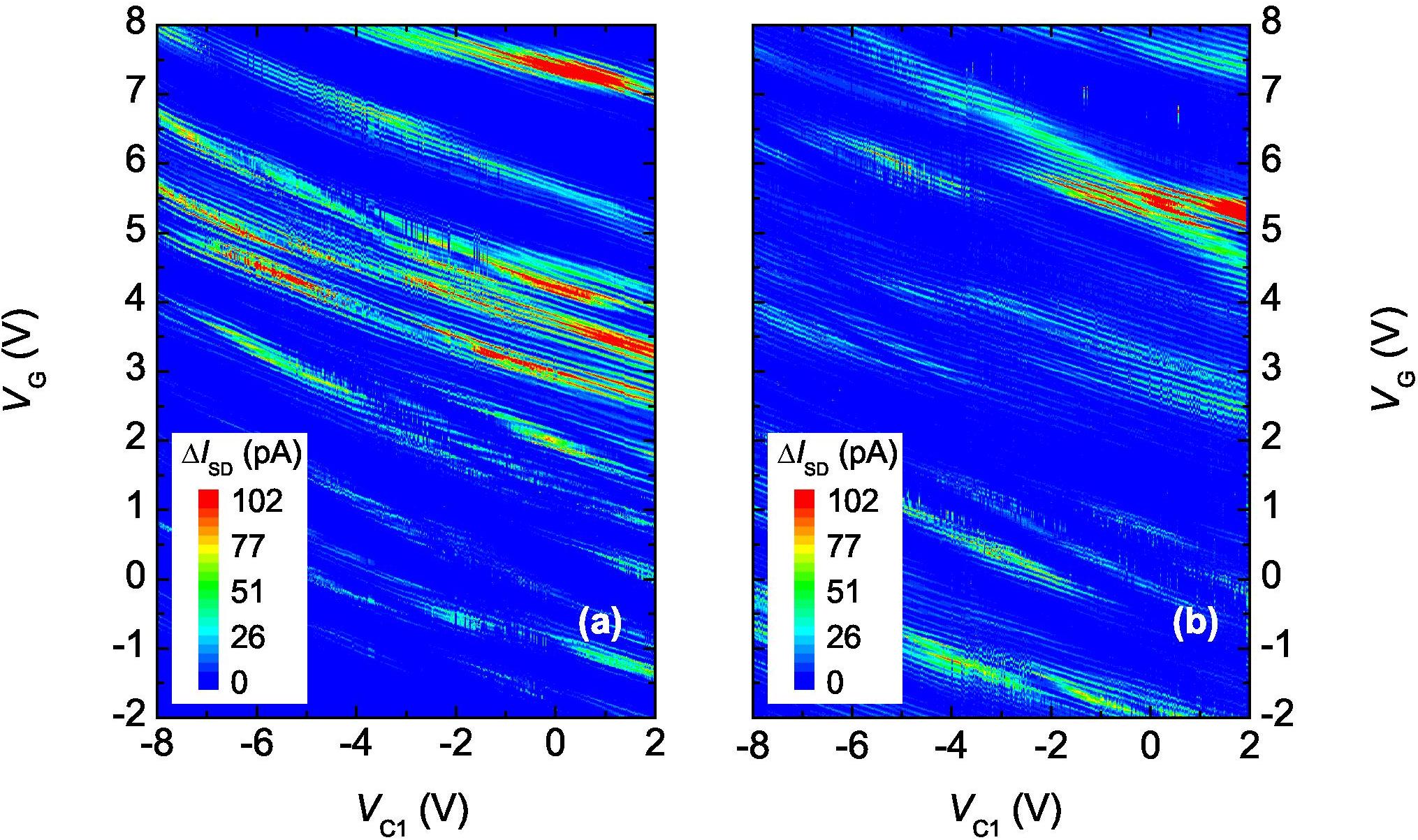}
\end{center}
\caption{\label{fig:figure6} Gate dependencies following different cooldowns both with a source-drain bias $V_{\textup{\tiny{SD}}}$= 3\,mV.}
\end{figure}

\subsection{Localization}

In doped semiconductors, the metallic phase is usually reached when the doping concentration exceeds the Mott critical limit $n_{\textup{\tiny{C}}}$, i.e. when the impurity band merges into the conduction band. In the case of phosphorous-doped bulk silicon, $n_{\textup{\tiny{C}}}\sim10^{18}$\,cm$^{-3}$ but, because of disorder, significant Lifshitz tails in the density of state (DOS) remain up to about 4$\times10^{19}$\,cm$^{-3}$ as shown by Altermatt \textit{et al.} \cite{Altermatt}. Therefore, localized states are still present above $n_{\textup{\tiny{C}}}$ but, because of their relative small number compared with extended states, they barely affect the conductivity that remains metallic. However, in reduced dimensions and, in particular, in quantum dots, the confinement increases electron-electron interaction and the presence of interfaces (especially non-(100) surfaces that are known to possess a high state density) play a significant role in the electron localization at the edge of the structure. In this case, Abrahams' scaling arguments and the concept of metal-insulator transition breaks down \cite{Abrahams}.

In our device, oxidation is used in order to reduce random telegraph signals and improve the noise performance. Nevertheless, this also redistributes phosphorous dopants towards the sidewalls of the structure, as demonstrated by the parabolic dopant profile obtained from secondary ion mass spectrometry (SIMS) experiments in similar devices  \cite{SIMS}. The effective dopant concentration is then decreased at the center of the island. Segregation, a process highly dependent on oxidation conditions \cite{Lau}, may also happen and dielectric screening at the interface may contribute to the localization at the sidewalls, in a manner that is dependent on processing conditions. 

The presence of impurity traps in a highly doped silicon device is thus not unusual. The existence of localized states was indeed demonstrated in microwave measurements and the temperature dependence of the conductivity \cite{Hasko}. Some weakly bound traps may be found inside the SET island. Still, most of the traps are expected to be found at the periphery of the SET island for the reasons given previously.
It should be noted that localization effects barely affect the largest area of the device, including contact leads that keep their metallic behaviors.

\section{Detector properties}

\subsection{General characteristics}

COs are clearly visible in the gate dependency diagram (Fig. B1) as lines with a slope $\textup{d}V_{\textup{\tiny{G}}}/\textup{d}V_{\textup{\tiny{C1}}} \sim$0.25$\pm 0.01$. Deviations from the linear dependence are noticeable for the most negative IDQD gate values ($V_{\textup{\tiny{C1}}} < -4$\,V). In these regions, the leakage current is still relatively small as compared with $I_{\textup{\tiny{SD}}}$ but non-linear effects in the structure may be induced at high voltages through a dependence of the capacitance values on gate voltages. 
The period of the oscillations $\Delta V_{\textup{\tiny{G}}}$ remains close to 96\,mV with no noticeable variation in gate voltage, and it is barely affected by thermal cycles. 
However, in all measured devices, the CO amplitude is strongly modulated by the conductivity background. The interplay between the COs and the conductivity background due to impurities is especially obvious when measuring the variation of the differential conductance with source-drain bias and observing the shape of the Coulomb diamonds (Fig. 5a). The latter are generally varying in dimensions and, in some cases, may appear as a convolution of diamonds of different sizes. 

\begin{figure}
\begin{center}
\includegraphics[width=85mm,bb=0 0 300 200]{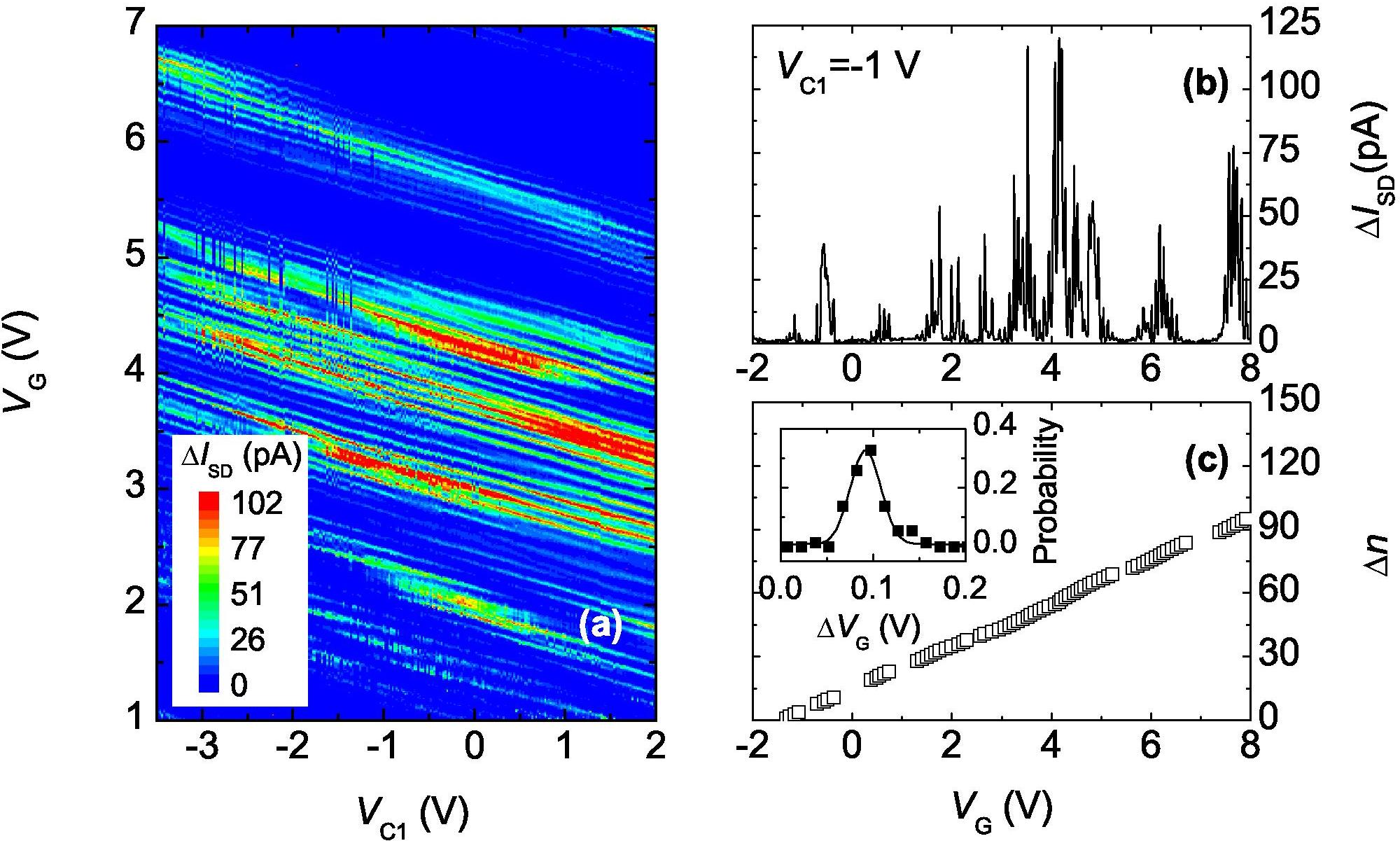}
\end{center}
\caption{\label{fig:figure7} a) Variation of $I_{\textup{\tiny{SD}}}$ with $V_{\textup{\tiny{G}}}$ and $V_{\textup{\tiny{C1}}}$.  b) SET gate oscillations for $V_{\textup{\tiny{C1}}} = -1$\,V. c) Coulomb peak positions in gate voltage for $V_{\textup{\tiny{C1}}} = 2$\,V showing a clear periodicity with a relative small Gaussian dispersion (inset). Due to the increasing gate leakage current for $V_{\textup{\tiny{G}}} < -2$\,V, the SET could not be depleted and $\Delta n$ represents the relative number of tunneling events in the SET island.}
\end{figure}

\begin{figure}
\begin{center}
\includegraphics[width=85mm,bb=0 0 300 200]{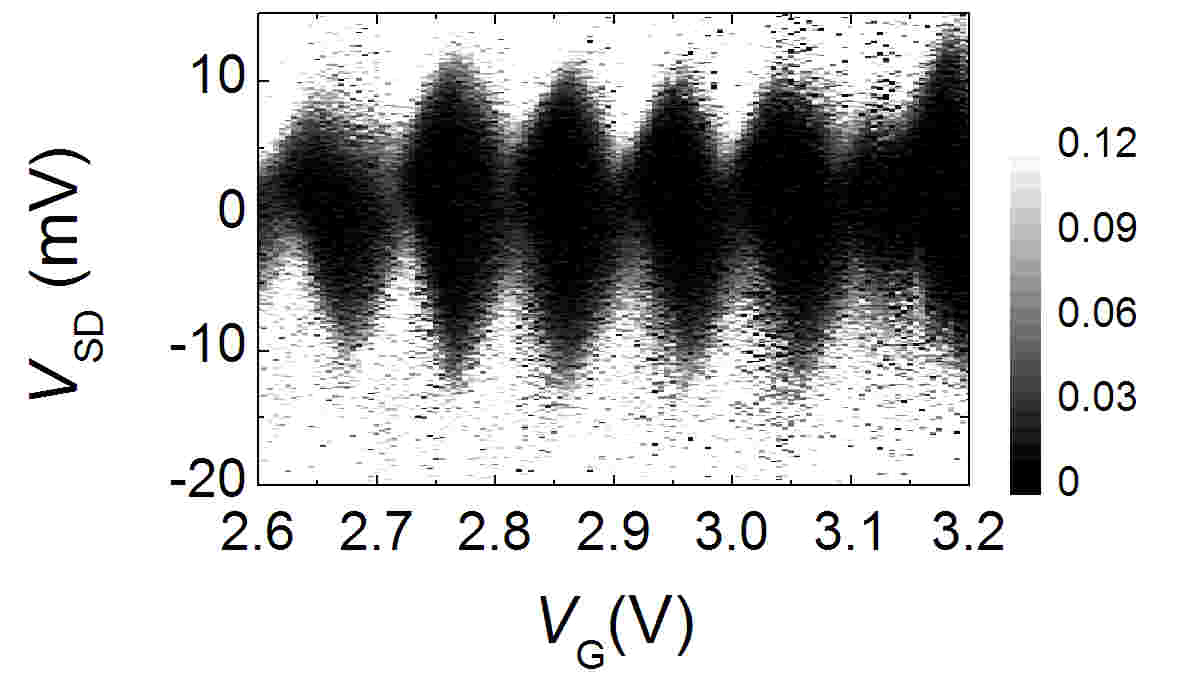}
\end{center}
\caption{\label{fig:figure8} Coulomb diamonds in undisturbed region for $V_{\textup{\tiny{C1}}} = -3$\,V.}
\end{figure}

Still, such structures are not observed everywhere and, in regions of gate voltages that are weakly affected by impurities, diamonds have a more regular shape, similarly to what is expected for metallic devices (Fig. B2). In such a region where the detector is behaving as a metallic-like island, capacitance approximation may be used and the information on the values of the capacitance between the different element of the structures as well as an approximation to the dot dimension are possible.

The slope of the Coulomb peak position in the gate dependency diagram is then given by a capacitance ratio $\sim C_{\textup{\tiny{C}}} / C_{\textup{\tiny{G}}}$ where $C_{\textup{\tiny{C}}}$ is the IDQD side gate to SET island capacitance  and $C_{\textup{\tiny{G}}}$ the SET gate to SET island capacitance (Fig. 5b).

For example, in the range 2.7\,V\,$<V_{\textup{\tiny{G}}}<$3.1\,V we obtain $C_{\textup{\tiny{G}}}= 1.7\pm 0.1$\,aF and from the values of the level arms we find $C_{\textup{\tiny{D}}}= 8.3\pm 0.7$\,aF and $C_{\Sigma}= 20.6\pm 4.1$\,aF respectively for the gate, drain and total SET dot capacitance. This gives a charging energy $E_{\textup{\tiny{C}}}\sim 4.0\pm 0.7\,$meV and a dot size of 32$\pm$ 6\,nm using the self capacitance for a metallic sphere. The difference in diameter values between the one determined by SEM imaging (60 nm) and the one determined by electrical characterization may be caused by an enhancement of background charge-induced confinement \cite{confinement}.

\subsection{Variable size quantum dot}

As previously seen, the conductivity in the device is influenced by the presence of localized states, so that a correct determination of the dot diameter is, in general, difficult, in common with most semiconductor devices of this type. In some region of gate voltages, the appearance of convoluted diamonds suggest the existence of parallel conduction channels in some devices \cite{Kodera} (Fig. 5a). These parallel transport mechanisms may lead, in some cases, to the suppression of the SET current over a large range of gate voltage. It is strongly affected by thermal cycles, suggesting that conduction involves bound states in the SET island (Fig. B3b) rather through a physically defined dot in the leads, barrier or inside the SET island. Following the discussion in Appendix A, it seems reasonable to approximate the quantum dot as a metallic-like sphere surrounded by a region of stronger localization. Within this region, the electron redistribution may happen due to a change in temperature, such as during thermal cycling. At low temperatures, electron activation from traps is strongly reduced. However, the strength of both disorder and electron interaction may be sufficient to induce hopping between sites and to allow a modification of the electrostatic environment following a change in the gate voltages. Indeed, the dimensionless parameter $r_{\textup{s}}$ \cite{rs} that characterize the relative strength of the electron interaction is about 1 in our device. This indicates that both electron-electron interactions and disorder play an important role.

A parallel conduction mechanism or its suppression may be explained by a change in the trap occupancy inducing a modification of the confining potential. It will appear in regions where the donor binding energy allow hopping between localized sites to happen, so, most likely, at the periphery of the the SET island. When all surrounding traps are occupied, the dot size is electrostatically reduced due to Coulomb repulsion and conduction via edge states is forbidden due to the absence of vacant sites. The Coulomb diamonds may then appear larger than expected. On the contrary, if edge states are all empty then the effective dot size appears larger and the blockade can be partly lifted due to electron hopping via edge states. As a consequence, we should observed a reduction of the CO period when the parallel conduction mechanism is increased. This effect is observed experimentally and provides a method for extracting the real size of the dot as well as the size of the localizing region.

In a region affected by trapping at the edge states but where Coulomb peaks are still clear and distinct, the largest Coulomb diamonds provide an estimate for the minimum effective dot size (all traps filled) whereas the smallest diamonds give the maximum dot size (all traps empty). From this, we obtain a diameter of 58\,nm with 14\,nm extent for the localizing region. The presence of traps at the edge of the structure, thus provides an effective mechanism for dot compression.

\begin{figure}
\begin{center}
\includegraphics[width=85mm,bb=0 0 300 200]{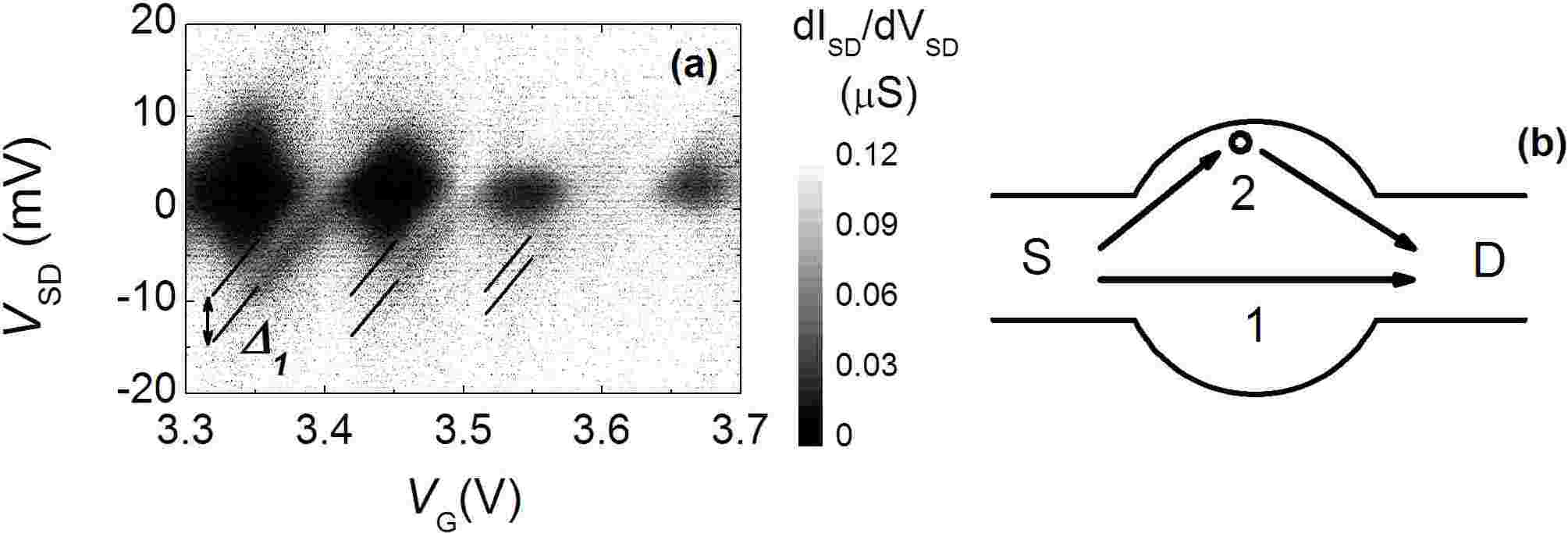}
\end{center}
\caption{\label{fig:figure9} a) Excited states. Contrast had to be enhanced to obtain a measurable energy value for the smallest diamond). b) Direct electron tunneling in the SET dot (1) between the source (S) and drain (D) contacts and possible parallel mechanism involving localized states at the dot periphery (2).}
\end{figure}

The quantum dot compression is more directly demonstrated by the change of the excited state energy with gate voltage, as the mean on particle level spacing ${\it{\Delta}}_1$ is inversely proportional to the dot size (Fig. B3a).
Indeed, this observation discards the possibility that the change in the shape of the Coulomb diamonds could be due to an increase of conductivity due to a modification of the tunneling barrier profile and, so of the electron tunneling rates. By neglecting the non-uniformity in the doping concentration, ${\it{\Delta}}_1$ can be estimated by equating the spatial extension of the ground state to the SET dot radius $R$ for a 2D isotropic harmonic oscillator confinement \cite{Tilke}:

\begin{eqnarray}\label{eqn:equation1}
 \frac{1} {2} m^{\star} {\omega_0}^2 R^2=\hbar \omega_0={\it{\Delta}}_1
\end{eqnarray}

where $m^{\star} = 0.19 m_{\textup{e-}}$ the transverse mass of electron in silicon and $\hbar \omega_0$ the quantum dot confinement energy.

For example, in the range $3.3 < V_{\textup{\tiny{G}}} < 3.7$ V, we obtain $E_{\textup{\tiny{C}}}\sim$ 4.4, 3.2 and 2.1 meV and ${\it{\Delta}}_1 \sim$ 2.1, 1.7 and 1.1 meV from which we obtain a dot size of about 30, 39 and 58 nm respectively for $V_{\textup{\tiny{G}}} =$ 3.35, 3.45 and 3.55 V. This is in good agreement with the absence of a depletion at $V_{\textup{\tiny{G}}} =$ 3.55 V and a depletion width of 14 nm for $V_{\textup{\tiny{G}}} =$ 3.35 V.

Finally, it is interesting to notice that the extrapolation of the charging energy in the region where Coulomb diamonds disappear (conduction via edge states) lead to a value $\sim 1.0$ meV similar to the excited states energy for a dot size of \,58 nm.

\section{Tunneling and trap assisted cotunneling}

In regions of gate voltages where the IDQD lines are not present and the conductivity not affected by the trap occupancy (diamond A in Fig. C1b), the SET charging energy can equivalently be calculated by measuring the width of a Coulomb diamond along $V_{\textup{\tiny{G}}}$ or along $V_{\textup{\tiny{SD}}}$ and converting the corresponding voltages into energies by using the value of the level arm respectively for the SET gate, e.g. $\alpha_{\textup{\tiny{G}}} = C_{\textup{\tiny{G}}} / C_{\Sigma}$ or the SET drain e.g. $\alpha_{\textup{\tiny{D}}}=C_{\textup{\tiny{D}}} / C_{\Sigma}$. We find $E_{\textup{\tiny{C}}} \sim 4.4$\,meV using $\alpha_{\textup{\tiny{G}}}$ and 4.2 meV using $\alpha_{\textup{\tiny{D}}}$.

For the diamond A, when the SET dot is blocked for conduction, the source-drain current follows $I_{\textup{\tiny{SD}}} \propto {V_{\textup{\tiny{SD}}}}^{\beta}$ with $\beta=3$ as expected for single electron tunneling through a two-tunnel junction because of cotunneling effect \cite{Tunneling}. On the other hand, when a large number of traps at the edge of the SET are occupied (peak B in Fig. C1b), we found $\beta=9$ and diamond widths along $V_{\textup{\tiny{SD}}}$ that extend well over the expected charging energy. The periodicity along $V_{\textup{\tiny{G}}}$ is also locally lost at zero source-drain bias.  

Although we observe a small variation in the values of capacitances and charging energy from peak to peak, a possible variation in the mean one-particle level spacing $E_1$ is not sufficient to explain the previous observation. However, a correct estimate of the charging energy and period could be obtained if one consider an anomalous suppression of cotunneling at the edge of the diamonds \cite{Luo}. As previously discussed, when all traps are filled, cotunneling via traps at the periphery is suppressed because vacant states are not available and Coulomb repulsion too strong for electron to tunnel directly via the SET island. 

From the expected value for the charging energy, one can estimate that inelastic cotunneling is associated with an energy difference of 1.5\,meV, a value close to the mean single particle level spacing estimated in the same region ($\sim 1.8$\,meV). Because electron motion in the IDQD is associated with tunneling via traps at the edge of the SET, this energy value has to be compared to the value of the charging energy at the shift region.

\begin{figure}
\begin{center}
\includegraphics[width=85mm,bb=0 0 300 200]{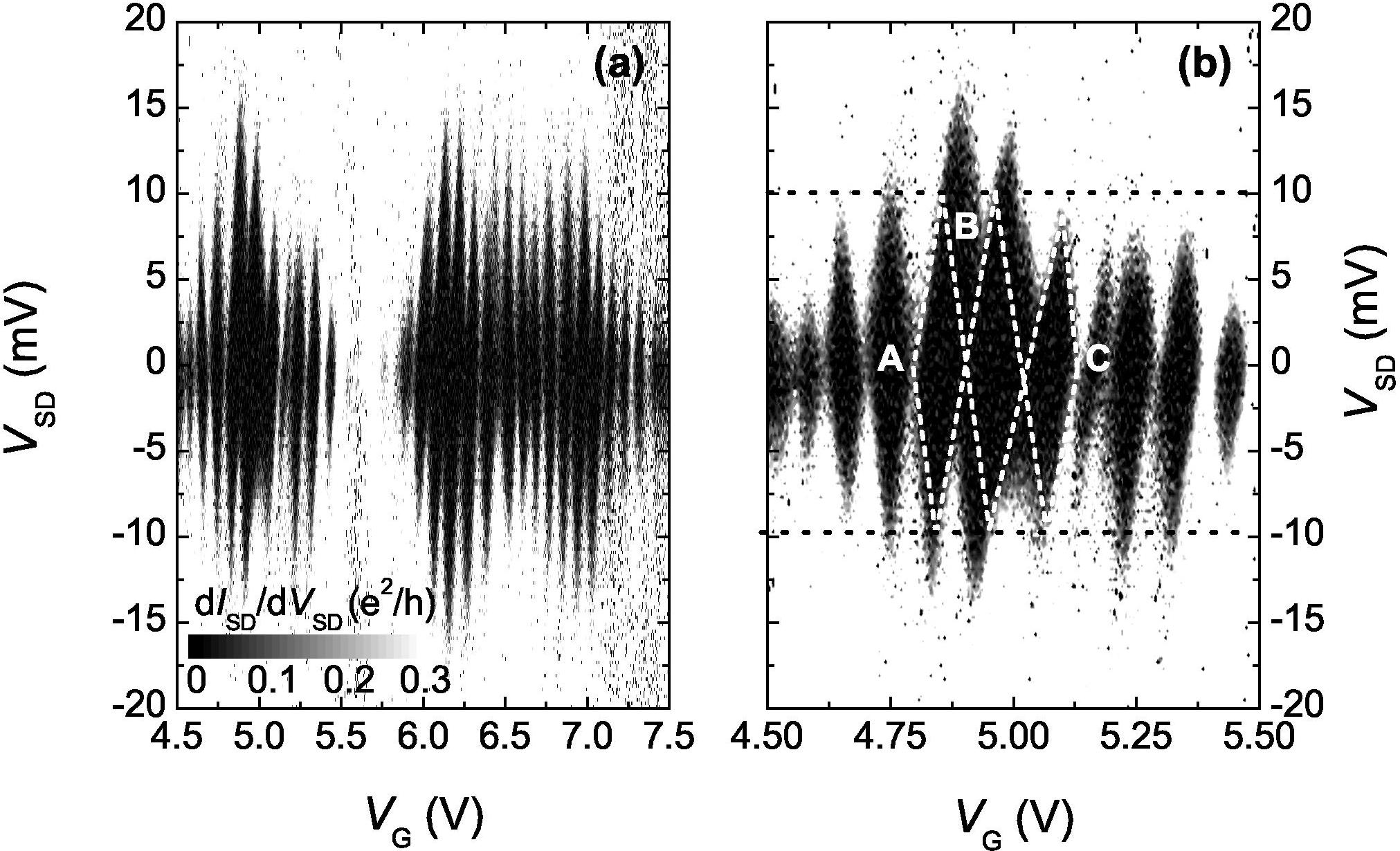}
\end{center}
\caption{\label{fig:figure10} a) Differential conductance d$I_{\textup{\tiny{SD}}}/$d$V_{\textup{\tiny{SD}}}$ versus $V_{\textup{\tiny{G}}}$ and $V_{\textup{\tiny{SD}}}$ at $V_{\textup{\tiny{C1}}}=0.5$\,V. b) Close-up view of Coulomb diamonds structure. In some cases, the first order tunneling is strongly suppressed (B) compared with expected diamonds (A). (C) shows a charging by a single impurity in the SET island.}
\end{figure}

To this end, we have used of the second IDQD gate that was previously grounded. The SET current is first mapped as a function of $V_{\textup{\tiny{G}}}$ and $V_{\textup{\tiny{C1}}}$ with $V_{\textup{\tiny{C2}}} = 0$ to choose a low noise region where a well defined shift is present and determine the IDQD-SET lines crossing point ${V_{\textup{\tiny{C1}}}}^{\star} = -6.3\,$ V and ${V_{\textup{\tiny{G}}}}^{\star} = 3.76\,$ V (Fig. C2a). In this region, the mean energy level spacing is about 1.1\,meV. In this experimental configuration, the variation of $I_{\textup{\tiny{SD}}}$ with $V_{\textup{\tiny{G}}}$ and $V_{\textup{\tiny{C2}}}$ for $V_{\textup{\tiny{C1}}} = {V_{\textup{\tiny{C1}}}}^{\star}$ has similar behavior as in figure C2a, and a clear shift centered on $V_{\textup{\tiny{C2}}} = 0$ is obtained. Because of its position, away from the SET island, $V_{\textup{\tiny{C2}}}$ can be used to detune the IDQD states from the degeneracy point without significantly affecting the SET at low biases. Coulomb diamonds are then obtained by varying $V_{\textup{\tiny{C2}}}$ across 0\,V  with $V_{\textup{\tiny{G}}} = {V_{\textup{\tiny{G}}}}^{\star}$ and $V_{\textup{\tiny{C1}}} = {V_{\textup{\tiny{C1}}}}^{\star}$ (Fig. C2b). 

The pattern of Coulomb diamonds is formed by a central diamond (I) with a charging energy of 2.1\,meV (between the two IDQD states), with two large side diamonds (II). These are followed by usual Coulomb diamonds (III) that are associated with the direct influence of $V_{\textup{\tiny{C2}}}$ on the SET ( $V_{\textup{\tiny{C2}}} > 0.5\,$V). The asymmetric shape of the diamonds (II) may be attributed to a difference in the charge polarization and the charge tunneling rates \cite{Tager}. At the IDQD-SET line crossing, the first order tunneling is suppressed and the conductivity is limited by higher orders of tunneling as explained in the previous sections. In figure C2b, the onset for conductivity gives an indication on the energy scale involved in the inelastic cotunneling process. Its value is closed to 1.0\,meV and, as expected, similar to the mean single particle energy spacing estimated for a 60\,nm diameter quantum dot (1.1\,meV).

These results, together with the observation of excited states in the device and the understanding of internal dynamics in the IDQD-SET-trap system, are consistent with the fact that the inelastic cotunneling via edge states is responsible for the observation of IDQD lines and, under appropriate coupling, the presence of shifts at specific value of gate voltages. The activation energy for the traps at the edge is varying between 1 and 2\,meV typically.

\begin{figure}
\begin{center}
\includegraphics[width=85mm,bb=0 0 300 200]{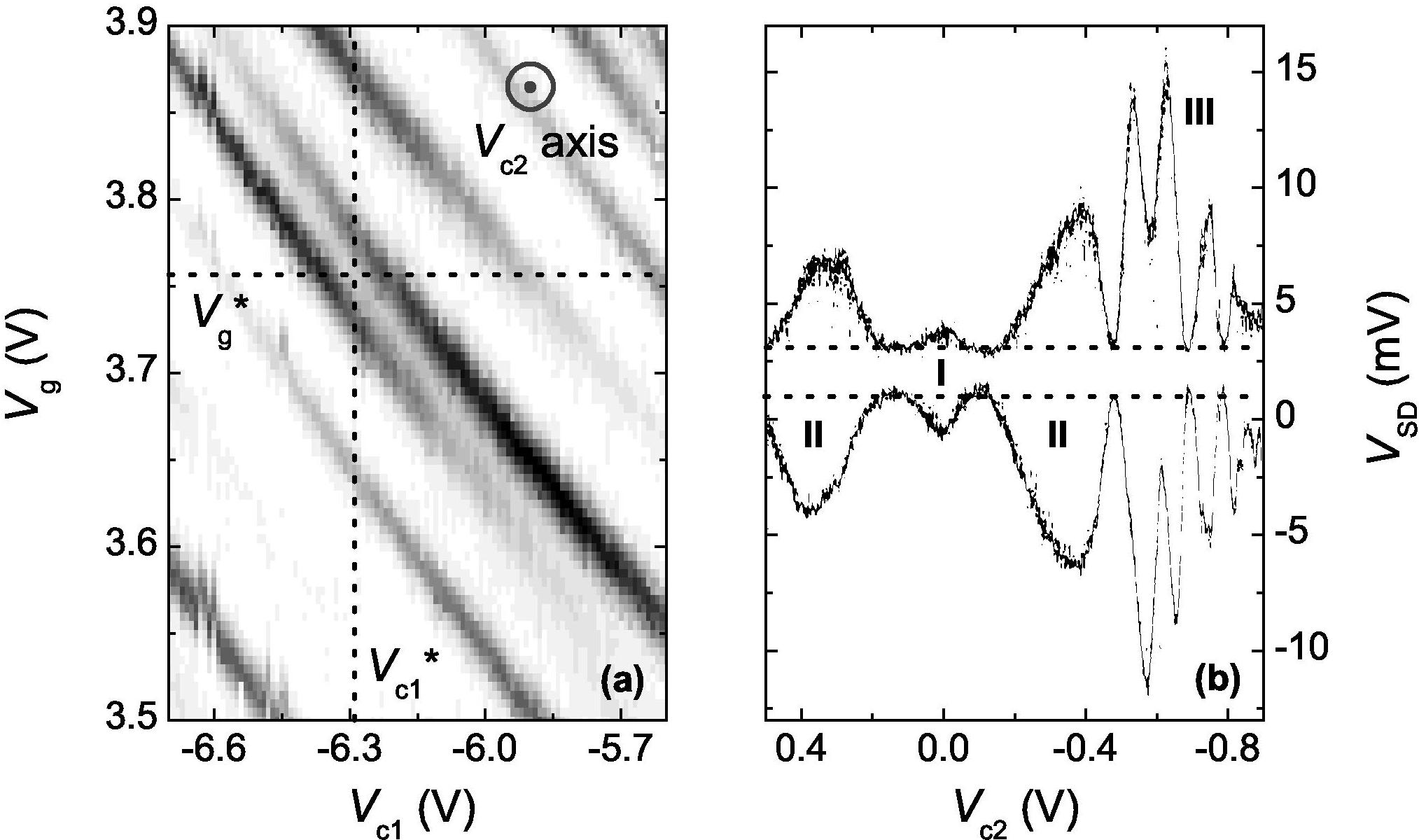}
\end{center}
\caption{\label{fig:figure11} a) Example of a Coulomb peak shift for $V_{\textup{\tiny{C2}}} = 0$. b) Contour plot at $I_{\textup{\tiny{SD}}}=I_{\textup{\tiny{SD}}} \left ({V_{\textup{\tiny{g}}}}^{\star}, {V_{\textup{\tiny{C1}}}}^{\star} \right)$ showing coulomb diamonds as a function of $V_{\textup{\tiny{C2}}}$.}
\end{figure}


\begin{thebibliography}{41}
\bibitem{Image} We consider an IDQD with permittivity $\epsilon_1$, a SET dot with the same permittivity that is separated from it by a distance $d$ and a medium of permittivity $\epsilon_2$. A charge $q$ located in the IDQD is then electrically equivalent to a charge $q'=\left( \epsilon_0 /\epsilon_1 \right) \left( \epsilon_2-\epsilon_1 \right) / \left( \epsilon_2+\epsilon_1 \right)$ in the SET, where $\epsilon_0$ is the permittivity of vacuum. By taking $\epsilon_1=11.7$ and $\epsilon_2=1$, we obtain $q'\sim 0.07 q$. Thus a variation of the IDQD charge by $q$ leads to a potential variation $\delta V_{\textup{\tiny{G}}}$ at the edge of the SET dot such as $\delta V_{\textup{\tiny{G}}}= q'/ \left( 4 \pi \epsilon_0 d \right) \sim 4.1$ mV or 4.3$\%$ of the CO period $\Delta V_{\textup{\tiny{G}}}= 96$ mV.
\bibitem{rs} The dimensionless interaction parameter $r_s$ usually characterizes the relative strength of Coulomb interaction in a structure. In a 2D quantum dot, it is related to the electron density $n_s$ by $r_s=\left(\pi n_s\right)^{-1/2}/a_0$ where $a_0={\hbar}^2 \epsilon/ \left(m e^2 \right)$ is the effective Bohr radius and, $m$ and $\epsilon$ are respectively the electron effective mass and the dielectric constant in silicon.
\end{thebibliography}
\end{document}